\documentclass[preprint,aps,epsfig]{revtex4}

\usepackage{psfig}

\def\lsim{\raise0.3ex\hbox{$<$\kern-0.75em\raise-1.1ex\hbox{$\sim$}}}
\def\gsim{\raise0.3ex\hbox{$>$\kern-0.75em\raise-1.1ex\hbox{$\sim$}}}

\newcommand{\be}{\begin{equation}}
\newcommand{\ee}{\end{equation}}

\def\beq{\begin{equation}}
\def\eeq{\end{equation}}
\def\beqa{\begin{eqnarray}}
\def\eeqa{\end{eqnarray}}

\newcommand{\rr}{\mbox{\boldmath $r$}}

\newcommand{\rb}{\mbox{\boldmath $b$}}

\def\gappeq{\mathrel{\rlap {\raise.5ex\hbox{$>$}}
{\lower.5ex\hbox{$\sim$}}}}

\def\lappeq{\mathrel{\rlap{\raise.5ex\hbox{$<$}}
{\lower.5ex\hbox{$\sim$}}}}

\def\Toprel#1\over#2{\mathrel{\mathop{#2}\limits^{#1}}}

\newcommand{\rk}{\mbox{\boldmath $k$}}

\begin{document}

\title{Saturation physics at HERA and RHIC: An unified description} 
\author{ V.P. Gon\c{c}alves$^1$,  M.S. Kugeratski$^2$, M. V. T. Machado $^3$, and  F.S. Navarra$^2$}
\affiliation{$^1$High and Medium Energy Group (GAME), \\
Instituto de F\'{\i}sica e Matem\'atica,  Universidade
Federal de Pelotas\\
Caixa Postal 354, CEP 96010-900, Pelotas, RS, Brazil\\
$^2$Instituto de F\'{\i}sica, Universidade de S\~{a}o Paulo, 
C.P. 66318,  05315-970 S\~{a}o Paulo, SP, Brazil\\
$^{3}$  High Energy Physics Phenomenology Group, GFPAE,  IF-UFRGS \\
Caixa Postal 15051, CEP 91501-970, Porto Alegre, RS, Brazil
}
\begin{abstract}
One of the  frontiers of QCD which are intensely investigated in high energy experiments 
is the high energy (small $x$) regime, where we expect to observe the non-linear behavior of the theory. 
In this regime, the growth of the parton distribution should saturate, forming a  Color Glass Condensate (CGC).
In fact, signals of parton saturation have already
been observed both in  $e p$ deep inelastic scattering at HERA and in deuteron-gold 
collisions at RHIC. Currently, a global description of the existing experimental data   
is possible considering different phenomenological saturation models for the two 
processes within the CGC formalism. In this letter we analyze the universality of these dipole 
cross section parameterizations and verify that they are not able to describe the HERA and RHIC 
data simultaneously. We analyze possible improvements in the parameterizations and propose  a 
new parametrization for the forward dipole amplitude  which { allows} us to   describe quite 
well the small-$x$ $ep$ HERA data { on} $F_2$ structure function as well as the $dAu$ RHIC data 
{ on} charged hadron spectra.  It is  an important signature of the universality of the 
saturation physics.

\end{abstract}
\maketitle
\vspace{1cm}

In  the past few years much theoretical
effort has been devoted towards the understanding of the high energy limit of the strong 
interaction theory (for recent 
reviews see, e.g. \cite{iancu_raju,weig,JK}). In the high energy limit, perturbative Quantum Chromodynamics (pQCD) predicts that the 
small-$x$ gluons in a hadron wavefunction should form a Color Glass Condensate (CGC), which is 
 described by an infinite hierarchy of the coupled evolution equations for the correlators of 
Wilson lines \cite{VENUGOPALAN,BAL,CGC,WEIGERT}.  In the absence of correlations, the first 
equation in the Balitsky-JIMWLK hierarchy decouples and is then equivalent to the equation 
derived independently by Kovchegov within the dipole formalism \cite{KOVCHEGOV}.
The Color Class Condensate  is characterized by the limitation on the maximum
phase-space parton density that can be reached in the hadron
wavefunction (parton saturation), with the transition being
specified  by a typical scale, which is energy dependent and is
called saturation scale $Q_{\mathrm{s}}$ [$Q_{\mathrm{s}}^2 \propto A^{\alpha} x^{-\lambda}$]. Moreover, in the CGC formalism    
 the   dipole-target forward scattering amplitude  ${\cal{N}}$  for a given impact 
parameter $\rb$, which is directly related with the  two-point function of Wilson lines,   encodes all the
information about the hadronic scattering, and thus about the
non-linear and quantum effects in the hadron wave function. The
function ${\cal{N}}$ can be obtained by solving the Balitsky-JIMWLK  evolution
equation in the rapidity $Y\equiv \ln (1/x)$ \cite{solutions}. Its  main properties
 are: (a) for the interaction of a small dipole ($\rr
\ll 1/Q_{\mathrm{s}}$), ${\cal{N}}(\rr) \approx \rr^2$, implying  that
this system is weakly interacting; (b) for a large dipole
($\rr \gg 1/Q_{\mathrm{s}}$), the system is strongly absorbed and therefore 
${\cal{N}}(\rr) \approx 1$.  This property is associated  to the
large density of saturated gluons in the hadron wave function.  Another 
 remarkable feature of CGC formalism is that the dense, 
saturated system of partons to be formed in hadronic wave functions at high energy has 
universal properties, the same for all hadrons or nuclei. 

In the CGC formalism the   description of the observables is directly related to the behavior of  ${\cal{N}}$. For instance,  the $F_2$ structure function is probed in $ep(A)$ process and is given by $F_2^{p (A)}(x,Q^2) = (Q^2/4 \pi^2 \alpha_{\mathrm{em}})(\sigma_T^{\gamma^*p(A)} + \sigma_L^{\gamma^*p(A)})$, where \cite{dipole}
\begin{eqnarray}
\sigma_{L,T}^{\gamma^*p(A)}(x,Q^2) & = & \sum_f \int dz \,d^2\rr
|\Psi_{L,T}^{(f)}(z,\rr,Q^2)|^2  \sigma_{dip}(x,\rr) \nonumber  \\
& = &  \sum_f \int dz \,d^2\rr
|\Psi_{L,T}^{(f)}(z,\rr,Q^2)|^2 \,\times  2\, \int d^2 \rb \, {\cal{N}_F}(x,\rr,\rb)\,\,, 
\label{sigdip}
\end{eqnarray}
with ${\cal{N}_F}$ being the fundamental representation of the forward dipole amplitude, 
 $\rr$ defining  the relative transverse
separation of the pair (dipole) and $z$ $(1-z)$ the
longitudinal momentum fraction of the quark (antiquark). The
photon wave functions $\Psi_{L,T}$ are determined from light cone
perturbation theory (See e.g. Ref. \cite{PREDAZZI}). It is useful to assume that the impact parameter dependence of $\cal{N}_{F}$ can be factorized as 
${\cal{N}_{F }}(x,\rr,\rb) = {\cal{N}_{F}}(x,\rr) S(\rb)$, so that 
$\sigma_{dip}(x,\rr) = {\sigma_0} \,{\cal{N}_F}(x,\rr)$, with $\sigma_0$ being   a free 
parameter  related to the non-perturbative QCD physics. Similarly, the single-inclusive hadron production in hadron-hadron  processes is described in the CGC formalism  by \cite{dhj}
\begin{eqnarray}
x_F{d\sigma^{p p(A) \rightarrow h X} \over dx_F \, d^2 p_t \, d^2 \rb} &=& 
{1 \over (2\pi)^2}
\int_{x_F}^{1} dx_p \, {x_p\over x_F} \Bigg[
f_{q/p} (x_p,Q_f^2)~ {\cal{N}_F} \left({x_p\over x_F} p_t ,\rb\right)~ D_{h/q} 
\left({x_F\over  x_p}, Q_f^2\right) +  \nonumber \\
&&
f_{g/p} (x_p,Q_f^2)~ {\cal{N}_A} \left({x_p\over x_F} p_t , \rb\right) ~ 
D_{h/g} \left({x_F\over x_p}, Q_f^2\right)\Bigg]~,
\label{eq:final}
\end{eqnarray}
where $p_t$ and $x_F$ are the transverse momentum and the Feynman-$x$
of the produced hadron, respectively. The variable $x_p$ denotes the momentum
fraction of a projectile parton and $\rb$ is the impact parameter.
Moreover,  $f(x_p,Q_f^2)$ is the projectile parton
distribution functions  and $D(z, Q_f^2)$ the parton fragmentation
functions into hadrons. These quantities  evolve according to the 
DGLAP~\cite{dglap} evolution equations and respect the momentum
sum-rule. In Eq. (\ref{eq:final}), ${\cal{N}_F}(\rk,\rb)$  and  ${\cal{N}_A} (\rk,\rb)$ 
are the fundamental and adjoint representations of the  forward dipole amplitude in  
momentum space. The amplitudes  ${\cal{N}_{A,F}}(\rk,\rb)$ and ${\cal{N}_{A,F}}(\rr,\rb)$ are 
directly related by a Fourier transform.


 The search of signatures for the parton saturation effects has been an
active subject of research in the last years.
In particular, it has  been observed that the HERA data at small $x$ and low $Q^2$ can be
successfully described with the help of saturation models \cite{GBW,bgbk,kowtea,iancu_munier,kowmot,sapeta}. 
Moreover,  experimental results for the total  \cite{scaling},  diffractive  \cite{marquet} and inclusive charm cross sections \cite{prl,hqcr} present the property of  geometric scaling.  On the other 
hand, the  observed \cite{BRAHMSdata}
suppression of high $p_T$ hadron yields at forward rapidities in dAu collisions at RHIC 
{ had} its behavior  anticipated on the basis of CGC ideas \cite{cronin}. 
A current shortcoming of these analyzes comes from the  non-existence of an exact solution of 
the non-linear equation in the full kinematic range, which implies the construction of 
phenomenological models satisfying the asymptotic behavior which { is} under theoretical 
control.   Several models for the forward dipole cross section have been used in the literature 
in order to fit 
the HERA and RHIC data. In particular, the phenomenological models from Refs. 
\cite{GBW,bgbk,kowtea,iancu_munier} { have} been proposed in order to describe the HERA data, 
while those from Refs. \cite{dhj,kkt} { have} been able to describe the $dAu$ RHIC data.
An important aspect should be emphasized at this point. Although at HERA it is possible to probe
  values of $x$ two orders of
 magnitude smaller than at RHIC,   the saturation scales for these two scenarios are very similar
 due to the nuclear medium (See Fig. 1 in Ref. \cite{kgn1}).   Consequently, one can expect to be
 possible to cross 
relate these experiments in this respect and gain a clear understanding of the CGC in high 
energy experiments.  There are several similarities among  the phenomenological  models proposed in Refs.\cite{GBW,bgbk,kowtea,iancu_munier,dhj,kkt}. In particular, in these models the function  ${\cal{N}}$ has been modeled in terms of a simple Glauber-like formula
\begin{eqnarray}
{\cal{N}}(x, \rr) = 1 - \exp\left[ -\frac{1}{4} (\rr^2 Q_s^2(x))^{\gamma (x,\rr^2)} \right] \,\,,
\label{ngeral}
\end{eqnarray}
where $\gamma$ is the anomalous dimension of the target gluon distribution.
The main difference comes from the  predicted behavior for the anomalous dimension, which determines  the  transition from the non-linear to the extended geometric scaling regimes, as well as from the extended geometric scaling to the DGLAP regime.  A detailed comparison has been presented in Ref. \cite{kgn1}. 
 As the models from Refs. \cite{GBW,bgbk,iancu_munier} have been exhaustively discussed in the literature, in this letter we only present a brief review of the models proposed in Refs.   
\cite{dhj,kkt}. In the KKT model \cite{kkt}  the expression 
for the quark dipole-target forward scattering amplitude  is given by \cite{kkt}:
\beq\label{glauber2}
{\cal{N}_F}(\rr,x) \, = \, 1- \exp\left[-\frac{1}{4} \left(\rr^2
\, \bar{Q}_s^2\right)^{\gamma(Y,\rr^2)}\right].
\eeq
where  $\bar{Q}_s^2 = \frac{C_F}{N_c} \, Q_s^2$ and  the  anomalous dimension 
$\gamma(Y, \rr^2)$  is 
\beq\label{gamma}
\gamma(Y, \rr^2) \, = \, \frac{1}{2}\left(1+\frac{\xi 
(Y, \rr^2)}{\xi (Y,\rr^2) + \sqrt{2 \,\xi (Y, \rr^2)}+  7
\zeta(3)\, c} \right),
\eeq
with $c$ a free parameter ({ which was fixed in \cite{kkt} to $c=4$}) and 
\beq\label{xi}
\xi (Y, \rr^2) \, = \, \frac{\ln\left[1/( \rr^2 \, Q_{s0}^2 ) 
\right]}{(\lambda/2)(Y-Y_0)}\,.
\eeq
The authors assume that the saturation scale can be expressed by $
Q_s^2(Y)  = \Lambda^2 A^{1/3} \left(\frac{1}{x}\right)^{\lambda}$. 
 The form of the anomalous
dimension is inspired by the analytical solutions to the BFKL equation
\cite{BFKL}. Namely, in the limit $\rr\rightarrow 0$ with $Y$ fixed we 
recover the anomalous dimension in the double logarithmic
approximation $\gamma \approx 1 - \sqrt{1/(2 \, \xi)}$. In another
limit of large $Y$ with $\rr$ fixed, Eq. (\ref{gamma}) reduces to the
expression of the anomalous dimension near the saddle point in the
leading logarithmic approximation $\gamma \approx
\frac{1}{2} + \frac{\xi}{14 \, c \, \zeta (3)}$. Therefore Eq. (\ref{gamma}) 
mimicks the onset of the geometric scaling region \cite{iancu_munier,IANCUGEO}. 
In the calculations of  Ref. \cite{kkt} it is assumed that a 
 characteristic value of $\rr$ is $\rr
\approx 1/(2 \, k_T)$ where $k_T$ is the transverse momentum of the valence 
quark and $\gamma$ was approximated by 
$\gamma(Y, \rr^2) \approx \gamma(Y,1/(4 \, k_T^2))$. In the above expressions the 
parameters $\Lambda=0.6$~GeV and 
$\lambda=0.3$ are fixed by DIS data \cite{GBW}. Moreover, the authors assume $Y_0 = 0.6$. 
The initial saturation scale used
in (\ref{xi}) is defined by $Q_{s0}^2=Q_s^2(Y_0)$ with $Y_0$ being  the
lowest value of rapidity at which the low-$x$ quantum evolution
effects are essential. As demonstrated in Ref. \cite{kkt} this parameterization is able to 
describe the $dAu$ RHIC data when the forward dipole cross section is convoluted with the 
respective fragmentation function and the parton distributions for the deuteron.
On the other hand, in Ref. \cite{dhj} another phenomenological saturation  model has been 
proposed in order to describe the $dAu$ RHIC data (hereafter denoted DHJ model). The basic 
modification with respect to the KKT model is the parameterization for the anomalous dimension 
which is now given by 
\beq\label{gamma_dhj}
\gamma(Y, \rr^2) \, = \gamma_s + \Delta \gamma (Y,\rr^2)
\eeq
where 
\beq
\Delta \gamma (Y,\rr^2) = (1 - \gamma_s) \frac{ |\log\frac{1}{\rr^2 Q_T^2}|}{\lambda Y +
 |\log\frac{1}{\rr^2 Q_T^2}| + d\sqrt{Y}}\,\,\,,
\eeq
with $Q_T = Q_s(Y)$ a typical hard scale in the process, $\lambda = 0.3$ and $d = 1.2$. Moreover, $\gamma_s = 0.63$ is the anomalous dimension for BFKL evolution with saturation boundary condition.  Similarly to the KKT model this model is 
able to describe the $dAu$ RHIC data.

\begin{figure}
\vspace*{1.5cm}
\centerline{
{\psfig{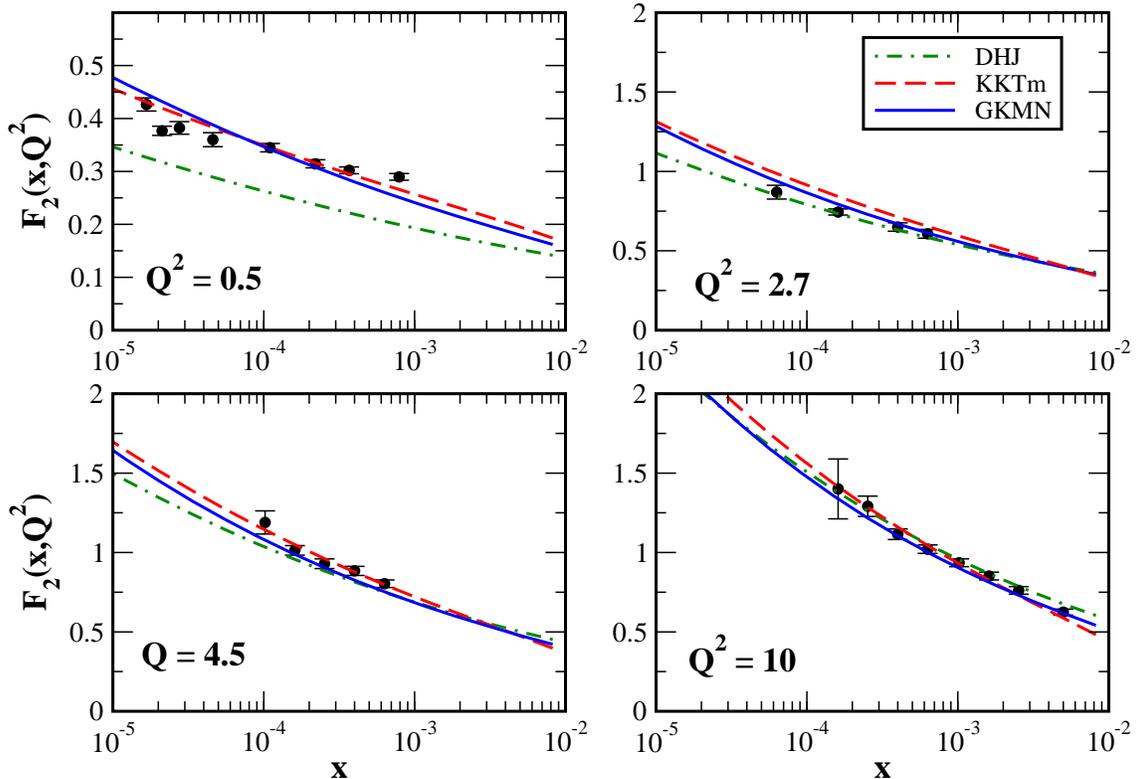}}}
\caption{The proton structure function at different values of the photon virtualities. Data from ZEUS.}
\label{fig1}
\end{figure}

\begin{figure}
\vspace*{1.5cm}
\centerline{
{\psfig{figure=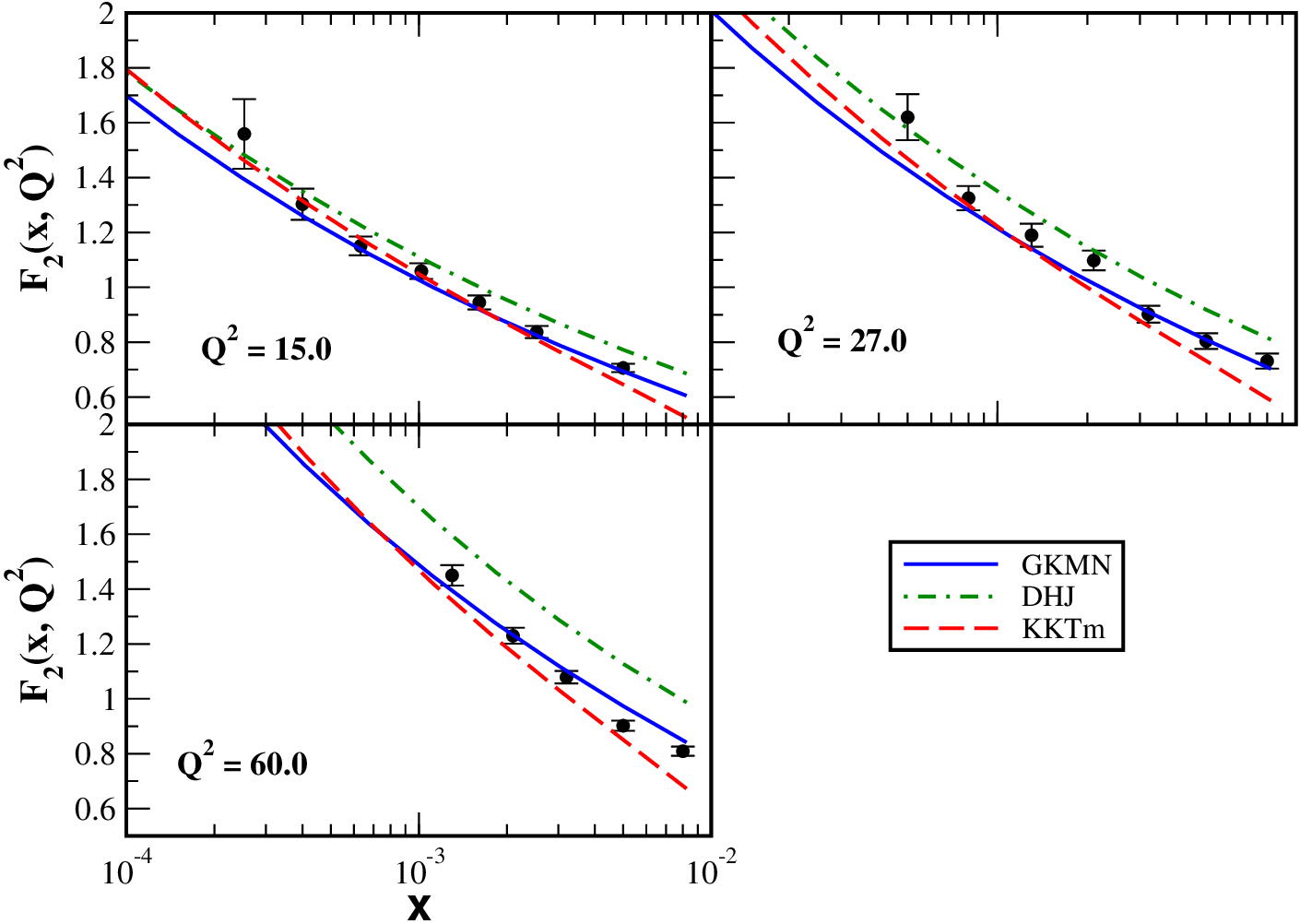,width=15.cm}}}
\caption{The proton structure function at different values of the photon virtualities. Data from ZEUS.}
\label{fig2}
\end{figure}

As already discussed in Ref. \cite{kgn1}, based on 
the universality of the hadronic 
wave function predicted by the CGC formalism,  we might expect that the KKT and DHJ 
parameterizations would also describe the HERA 
data on  proton structure functions  in the  kinematical 
region where the saturation effects should be present (small $x$ and low $Q^2$). However, this 
expectation fails when the KKT model  is applied, as verified in \cite{kgn1}. Here we extend 
{ the  analysis to}  the DHJ model without any modification of the parameters fitted at RHIC, 
only assuming $A =1$ and adjusting the  non-perturbative parameter $\sigma_0$, which defines the 
normalization, in order  to describe the $F_2$ experimental data at $Q^2 = 10$ GeV$^2$.   
In Figs. \ref{fig1} and \ref{fig2} we present the predictions of the DHJ model for the  proton 
structure function and compare with the ZEUS data \cite{zeus}. We can see that this 
parameterization fails for both small and large values of $Q^2$. Consequently, the current
 parameterizations of the forward dipole cross section which are constrained at RHIC are not 
able to describe the HERA data. An open question is if minimal modifications in these 
parameterizations allow to describe both { sets} of data.   
Following Ref. \cite{magnofl} we { consider} a modification of the KKT model assuming that 
the saturation momentum scale is given as in the GBW model, $Y_0 = 4.6$, $c = 0.2$ and that the 
typical scale in the computation of $\xi (Y, \rr^2) $ is the photon virtuality. Its predictions 
(KKTm lines) are presented in Figs. \ref{fig1} and \ref{fig2}. It is observed that  these
 modifications imply a quite good description of the HERA data. Similarly, as the $Q^2$ evolution of the $F_2$ data is not well described by the DHJ model it is possible to improve this model by the modification of the anomalous dimension. Here we propose to modify the 
DHJ model assuming now that $Q_T = Q_0 = 1.0$ GeV, {\it i. e.} that the typical scale is energy independent. It is important to emphasize that this modification preserves the main properties of the anomalous dimension proposed in \cite{dhj}. Basically, we still have that the anomalous dimension increases logarithmically with $p_T$ from $\gamma = \gamma_s$ to its asymptotic value $\gamma \approx 1$, while decreasing with $Y$ as $\Delta \gamma \approx 1/Y$ at very large rapidity. As shown in Figs. \ref{fig1} and 
\ref{fig2}, with this modification our predictions (GKMN lines)  agree with the experimental data.

\begin{figure}
\vspace*{1.5cm}
\centerline{
{\psfig{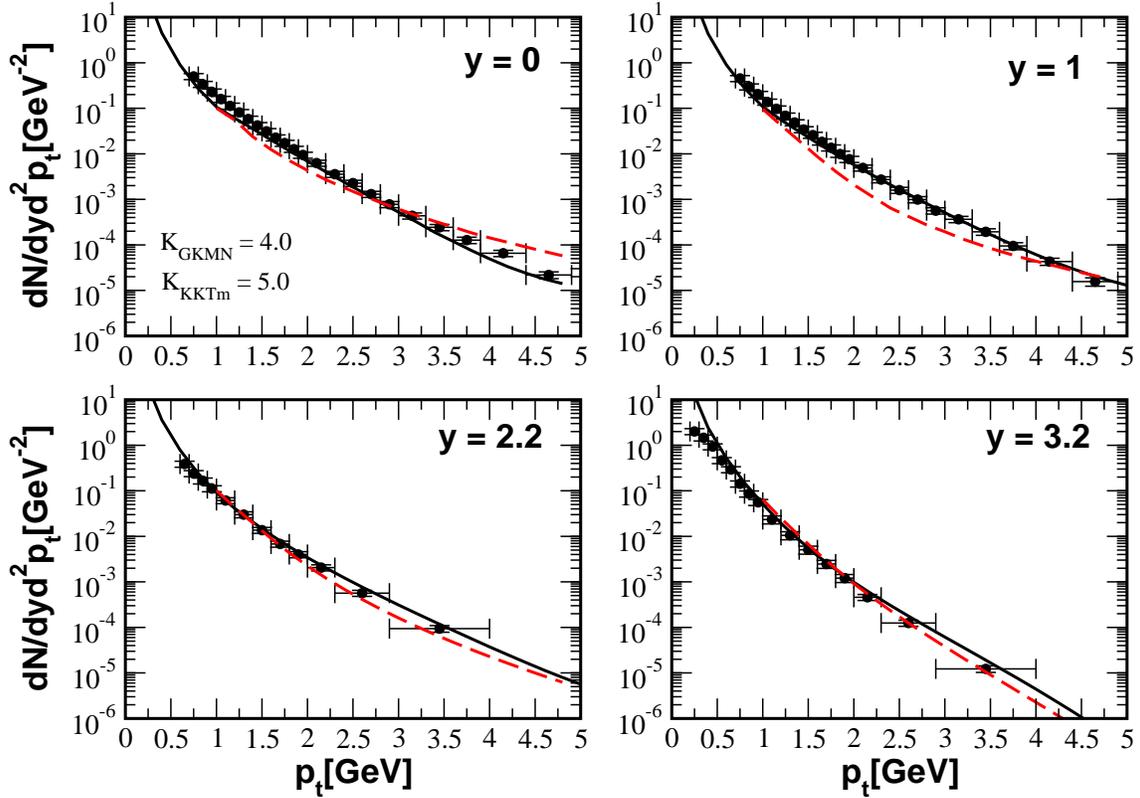}}}
\caption{Comparison of theory and BRAHMS data for minimum-bias $dAu$ collisions at RHIC energy. }
\label{fig3}
\end{figure}

The question which follows is { whether the RHIC data are still well reproduced} after these
modications. Following Ref. \cite{dhj} we have calculated the single inclusive hadron production
{ cross section} in $dAu$ collisions at different rapidities. We have used the CTEQ5L quark 
and gluon distributions \cite{cteq} and the LO KKP quark-hadron fragmentation functions  
\cite{kkp}. Our results are presented in Fig. \ref{fig3} and compared with the BRAHMS data 
\cite{BRAHMSdata}. The KKTm  and GKMN  predictions are represented by long-dashed and solid 
curves respectively. { As in Ref. \cite{dhj} we need  a $K$-factor} in our calculations, 
since it has been performed at leading order in $\alpha_s$. Although the normalization should be 
modifed by these corrections we expect that the shape of the momentum distributions should not  
change. { Our  values of $K$ have been determined so as to reproduce the 
data at $p_T = 1.0$ GeV and they depend on the parameterization adopted. For KKTm we find a
larger value of $K$ than for GKMN. Moreover,  while the KKTm parameterization fails to describe 
the full set of data, the GKMN one is able to reproduce the data quite well even at very small 
values of $p_T$.}
Consequently, the GKMN model is able to describe the $ep$ HERA and $dAu$ RHIC data in terms of 
an unique parameterization for the dipole scattering amplitude, which is based on the saturation 
physics.

Before presenting a summary of our main results, let us briefly discuss the basic properties of 
the resulting GKMN model (a more detailed analysis will be presented elsewhere).  In Fig. 4 (a) 
we present the forward dipole cross section as a function of the scaling variable $rQ_s$ for 
distinct parameterizations. As it can be seen the DHJ, KKTm and GKMN models have a similar 
behavior. The difference among  the models can be demonstrated studying the $Q^2$ behaviour 
of the  effective anomalous dimension, defined by 
$\gamma_{eff} = \frac{ d \ln {\cal{N}}(rQ_s,Y)}{d \ln (r^2 Q_s^2/4)}$ (See similar analyzes in 
Ref. \cite{magnofl}). In Fig. 4 (b) is shown $\gamma_{eff}$ as a  function of the virtuality 
$Q^2$, using the average dipole size as $r = 2/Q$. While the GBW model presents a fast 
convergence to the DGLAP anomalous dimension at large $Q^2$,  the IIM parameterization has a 
mild growth with  virtuality, converging to $\gamma \approx 0.85$ at large $Q^2$. The KKTm and 
IIM parameterizations are similar at large $Q^2$, but differ at small virtualities, with the KKTm one predicting a smaller value. On the other hand, the predictions of the DHJ and GKMN parameterizations are similar at small $Q^2$ and differ at large virtualities. In particular, we have a strict difference between these models in the intermediate range of virtualities, which can explain why the DHJ model does not describe the $Q^2$ evolution of the $F_2$ structure function.

As a summary, in this letter we have analyzed current parameterizations for the dipole 
scattering amplitude which are able to describe separately the $ep$ HERA and $dAu$ RHIC data.  
We have shown  that an unified description using these parameterizations
is not possible.  We have proposed a modification in the 
DHJ parameterization for the dipole scattering amplitude, based 
on  saturation physics, which allows to describe simultaneously the  $ep$ HERA and $dAu$ 
RHIC data. This result has been obtained   adjusting the normalizations of the dipole cross 
section and single inclusive hadron cross section and  assuming an energy independent typical 
scale, keeping all other original parameters. 
{A global least $\chi^2$ fit of data would change slightly the values of our parameters. 
This would be a fine tunning which is beyond the scope of this work. We rather prefer to keep 
the level of fitting accuracy of \cite{dhj} and emphasize  { the strategy to reconcile two 
different sets of data}.
}
Apart from this fine tunning, a more detailed theoretical study of the  proposed anomalous 
dimension is necessary. We postpone these improvements for a future publication. 
Finally, our  results demonstrate that an unified 
description of the experimental data which probes the high energy regime of QCD is possible. 
This is an important signature of the universality of the saturation physics. 



\begin{figure}[t]
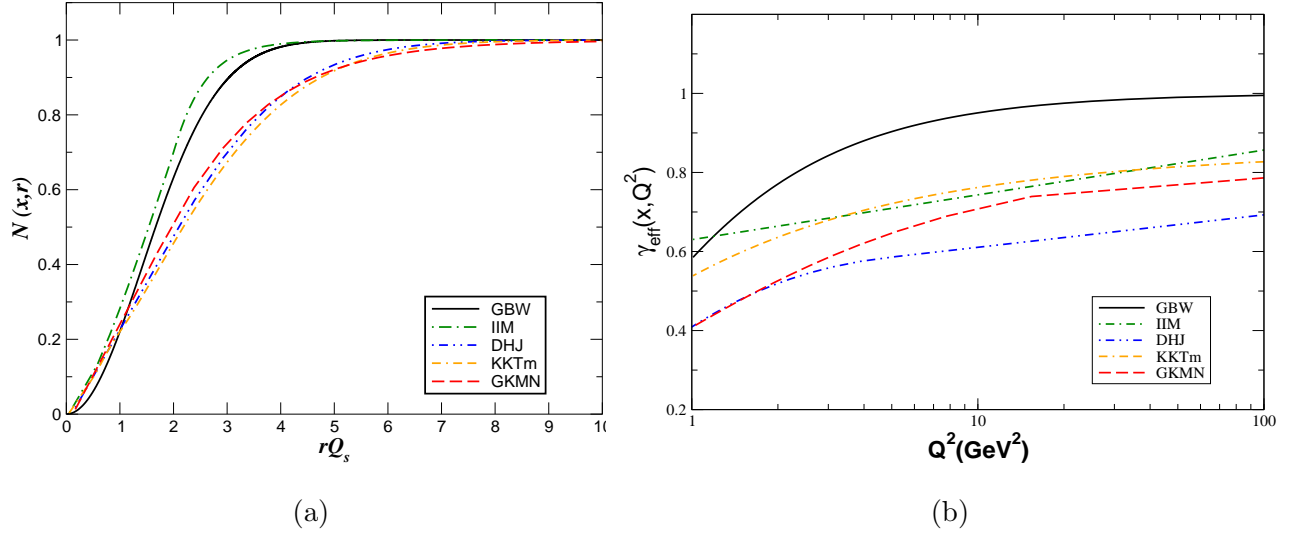

\begin{tabular}{cc}
{\psfig{figure=enerqs_gkmn2.eps,width=8.0cm}}  &
{\psfig{figure=gammaeff_q2.eps,width=8.7cm}}\\
(a) &  (b)
\end{tabular}
\caption{(a) Forward dipole cross section as a function of the scaling variable $rQ_{s}$. (b) The $Q^2$ behavior of the effective anomalous dimension at $x = 3 \times 10^{-4}$.}
\label{fig4}
\end{figure}

\begin{acknowledgments}
 VPG would like to thanks M. A. Betemps for  informative and  helpful discussions.   This work was  partially 
financed by the Brazilian funding
agencies CNPq, FAPESP and FAPERGS.
\end{acknowledgments}


\end{document}